\pdfoutput=1

\RequirePackage[l2tabu,orthodox]{nag}

\documentclass[11pt,letterpaper,reqno,oneside]{article}

\DeclareTextFontCommand{\emph}{\slshape}

\usepackage[T1]{fontenc}
\usepackage{lmodern}
\usepackage[utf8]{inputenc}

\usepackage[american,german,french,british]{babel}
\usepackage[activate={true,nocompatibility}]{microtype}

\usepackage{csquotes}

\usepackage[backend=biber,autolang=other,style=apa,maxcitenames=5,sortcites=false]{biblatex}
\DeclareLanguageMapping{british}{british-apa}
\DeclareLanguageMapping{american}{american-apa}
\DeclareLanguageMapping{german}{german-apa}
\DeclareLanguageMapping{french}{french-apa}
\DefineBibliographyExtras{french}{\restorecommand\mkbibnamefamily}
\usepackage{amsmath}
\usepackage{amssymb}
\usepackage{amsfonts}
\usepackage{amsthm}
\usepackage[retainorgcmds]{IEEEtrantools}
\usepackage{mathtools}
\usepackage{stmaryrd}
\usepackage{mleftright}
\usepackage{centernot}
\usepackage{graphicx}
\usepackage{tikz,pgfplots}
\pgfplotsset{compat=1.10}
\usetikzlibrary{shapes.multipart}
\usetikzlibrary{decorations.markings}
\usepackage{etoolbox}
\AtBeginEnvironment{tikzpicture}{\shorthandoff{;}}
\usepackage[titletoc,title]{appendix}

\usepackage{datetime}
\newdateformat{datestyle}{ {\THEDAY} \monthname[\THEMONTH] \THEYEAR }

\usepackage{enumerate}
\usepackage{enumitem}
\setlist[enumerate,1]{label=(\arabic*)}
\setlist[itemize,1]{label=--}
\setlist[itemize,2]{label=--}
\setlist[itemize,3]{label=--}
\setlist[itemize,4]{label=--}

\usepackage{xcolor}
\usepackage{verbatim}
\usepackage{gensymb}
\usepackage{rotating}
\usepackage{subcaption}
\usepackage{floatpag}
\floatpagestyle{plain}
\usepackage{marvosym}
\usepackage{hyphenat}
\theoremstyle{definition}
\newtheorem{proposition}{Proposition}%[section]
\newtheorem{lemma}{Lemma}%[section]
\newtheorem{corollary}{Corollary}%[section]
\newtheorem{remark}{Remark}%[section]
\newtheoremstyle{named}
	{\topsep}					% ABOVESPACE
	{\topsep}					% BELOWSPACE
	{}							% BODYFONT
	{0pt}						% INDENT (empty value is the same as 0pt)
	{\bfseries}					% HEADFONT
	{}							% HEADPUNCT
	{5pt plus 1pt minus 1pt}	% HEADSPACE
	{\thmnote{#3}}				% CUSTOM-HEAD-SPEC
\theoremstyle{named}
\newtheorem{namedthm}{}

\usepackage{xpatch}
\xpatchcmd{\proof}{\itshape}{\proofheadfont}{}{}
\newcommand{\proofheadfont}{\slshape}

\usepackage{hyperref}
\hypersetup{
	colorlinks=true,
	linkcolor=black,
	citecolor=black,
	filecolor=black,
	urlcolor=black,
}

\usepackage[nameinlink]{cleveref}
\crefname{equation}{equation}{equations}
\crefname{section}{section}{sections}
\crefname{subsection}{section}{sections}
\crefname{subsubsection}{section}{sections}
\crefname{appsec}{appendix}{appendices}
\crefname{supplsec}{supplemental appendix}{supplemental appendices}
\crefname{footnote}{footnote}{footnotes}
\crefname{figure}{figure}{figures}
\crefname{table}{table}{tables}
\crefname{theorem}{theorem}{theorems}
\crefname{proposition}{proposition}{propositions}
\crefname{lemma}{lemma}{lemmata}
\crefname{corollary}{corollary}{corollaries}
\crefname{remark}{remark}{remarks}
\crefname{observation}{observation}{observations}
\crefname{example}{example}{examples}
\crefname{fact}{fact}{facts}
\crefname{definition}{definition}{definitions}
\crefname{assumption}{assumption}{assumptions}
\crefname{exercise}{exercise}{exercises}
\crefname{notation}{notation}{notation}
\crefname{claim}{claim}{claims}
\crefname{conjecture}{conjecture}{conjectures}

\newcommand{\eps}{\varepsilon}

\newcommand{\dd}{\mathrm{d}}

\newcommand{\R}{\mathbf{R}}

\DeclarePairedDelimiter\abs{\lvert}{\rvert}

\makeatletter
\let\save@mathaccent\mathaccent
\newcommand*\if@single[3]{%
	\setbox0\hbox{${\mathaccent"0362{#1}}^H$}%
	\setbox2\hbox{${\mathaccent"0362{\kern0pt#1}}^H$}%
	\ifdim\ht0=\ht2 #3\else #2\fi
	}
\newcommand*\rel@kern[1]{\kern#1\dimexpr\macc@kerna}
\newcommand*\widebar[1]{\@ifnextchar^{{\wide@bar{#1}{0}}}{\wide@bar{#1}{1}}}
\newcommand*\wide@bar[2]{\if@single{#1}{\wide@bar@{#1}{#2}{1}}{\wide@bar@{#1}{#2}{2}}}
\newcommand*\wide@bar@[3]{%
	\begingroup
	\def\mathaccent##1##2{%
	  \let\mathaccent\save@mathaccent
	  \if#32 \let\macc@nucleus\first@char \fi
	  \setbox\z@\hbox{$\macc@style{\macc@nucleus}_{}$}%
	  \setbox\tw@\hbox{$\macc@style{\macc@nucleus}{}_{}$}%
	  \dimen@\wd\tw@
	  \advance\dimen@-\wd\z@
	  \divide\dimen@ 3
	  \@tempdima\wd\tw@
	  \advance\@tempdima-\scriptspace
	  \divide\@tempdima 10
	  \advance\dimen@-\@tempdima
	  \ifdim\dimen@>\z@ \dimen@0pt\fi
	  \rel@kern{0.6}\kern-\dimen@
	  \if#31
	    \overline{\rel@kern{-0.6}\kern\dimen@\macc@nucleus\rel@kern{0.4}\kern\dimen@}%
	    \advance\dimen@0.4\dimexpr\macc@kerna
	    \let\final@kern#2%
	    \ifdim\dimen@<\z@ \let\final@kern1\fi
	    \if\final@kern1 \kern-\dimen@\fi
	  \else
	    \overline{\rel@kern{-0.6}\kern\dimen@#1}%
	  \fi
	}%
	\macc@depth\@ne
	\let\math@bgroup\@empty \let\math@egroup\macc@set@skewchar
	\mathsurround\z@ \frozen@everymath{\mathgroup\macc@group\relax}%
	\macc@set@skewchar\relax
	\let\mathaccentV\macc@nested@a
	\if#31
	  \macc@nested@a\relax111{#1}%
	\else
	  \def\gobble@till@marker##1\endmarker{}%
	  \futurelet\first@char\gobble@till@marker#1\endmarker
	  \ifcat\noexpand\first@char A\else
	    \def\first@char{}%
	  \fi
	  \macc@nested@a\relax111{\first@char}%
	\fi
	\endgroup
}
\makeatother

\usepackage{atbegshi,picture}
\AtBeginShipoutNext{\AtBeginShipoutUpperLeft{%
	\put(\dimexpr(\paperwidth-\textwidth)/2\relax,-2.5cm\relax){\parbox[t]{\textwidth}{\footnotesize This paper was published in \emph{Mathematical Social Sciences} in 2020. The published version may be found at \href{https://doi.org/10.1016/j.mathsocsci.2020.07.002}{doi:10.1016/j.mathsocsci.2020.07.002}.
	This work is licensed under the \href{http://creativecommons.org/licenses/by-nc-nd/4.0/}{Creative Commons Attribution-NonCommercial-NoDerivatives 4.0 International License}.}}%
}}

\title{\scshape Strictly strategy-proof auctions%
\thanks{We are grateful to Yi-Chun Chen, Eddie Dekel, Jeff Ely, Piero Gottardi,
Matt Jackson, Bart Lipman, Alessandro Pavan, Marciano Siniscalchi, Asher Wolinsky
and two anonymous referees
for helpful comments. Part of this work was done while the first author was affiliated with the École des hautes études en sciences sociales (EHESS). The first author gratefully acknowledges financial support from the Ministère de l'enseignement supérieur et de la recherche (MESR) and from the European University Institute. The second author is grateful to Northwestern University for financial support in the form of a graduate fellowship.}}

\author{Matteo Escudé \\ 
European University Institute 
\and 
Ludvig Sinander \\ 
Northwestern University}

\date{25 July 2020}

\makeatletter
	\AtBeginDocument{ \hypersetup{ 
		pdftitle = {Strictly strategy-proof auctions}, 
		pdfauthor = {Matteo Escudé and Ludvig Sinander} 
		} }
\makeatother

\begin{document}

\maketitle

%%%%%%%%%%%%%%%%%%%%%%
%%%%%%%%%%%%%%%%%%%%%%
\begin{abstract}
	A strictly strategy-proof mechanism is one that asks agents to use \emph{strictly} dominant strategies. In the canonical one-dimensional mechanism design setting with private values, we show that strict strategy-proofness is equivalent to \emph{strict} monotonicity plus the envelope formula, echoing a well-known characterisation of (weak) strategy-proofness. A consequence is that strategy-proofness can be made strict by an arbitrarily small modification, so that strictness is `essentially for free'.
	%
	% \vspace{0.5em}
	%
	% \emph{Keywords:} Mechanism design, auction, strategy-proof, dominant strategy, robustness, virtual.
	%
\end{abstract}
%%%%%%%%%%%%%%%%%%%%%%
%%%%%%%%%%%%%%%%%%%%%%

%%%%%%%%%%%%%%%%%%%%%%
%%%%%%%%%%%%%%%%%%%%%%
\section{Introduction}
\label{sec:introduction}
%%%%%%%%%%%%%%%%%%%%%%
%%%%%%%%%%%%%%%%%%%%%%

Two popular notions of robustness in mechanism design are strategy-proofness and full implementation. The former requires that agents use weakly dominant strategies; the latter that \emph{every} equilibrium leads to the desired outcome.

In applications, it is common to have the one without the other. Consider the canonical auction setting with valuations identically and independently distributed on compact supports.
The first-price auction has a unique Bayes--Nash equilibrium implementing the efficient allocation,%
	\footnote{Under mild distributional assumptions. See e.g. \textcite{Lebrun2006}.}
but agents' strategies are not weakly dominant. The second-price auction has an efficient equilibrium in weakly dominant strategies, but also has many other, inefficient, equilibria.%
	\footnote{For example: agent 1 always bids strictly above the supports of all agents' valuations, and all other agents always bid zero.}

A sufficient condition for both kinds of robustness is \emph{strict} strategy-proofness: agents use \emph{strictly} dominant strategies. Weak strategy-proofness obviously follows, and full implementation follows because a game can have at most one strictly dominant strategy profile.
Strict strategy-proofness has the further appeal of strategic simplicity:
since each agent has a unique strictly dominant strategy,
she need not reason about her opponents' behaviour.

In this paper, we study strict strategy-proofness in the canonical one-dimensional mechanism design setting with private values and quasi-linear and strictly single-crossing preferences.
It is well-known that weak strategy-proofness is equivalent to monotonicity plus the envelope formula; we show in §\ref{sec:environment} that \emph{strict} strategy-proofness is equivalent \emph{strict} monotonicity plus the envelope formula.

In §\ref{sec:free}, we derive the implication that any weakly strategy-proof mechanism is \emph{virtually} strictly strategy-proof, meaning that it can be made strictly strategy-proof by an arbitrarily small modification. This can be viewed as a novel robustness property of weak strategy-proofness. It follows further that a principal designing a weakly strategy-proof mechanism can achieve \emph{strict} strategy-proofness at arbitrarily small cost.

%%%%%%%%%%%%%%%%%%%%%%%%%%%%
%%%%%%%%%%%%%%%%%%%%%%%%%%%%
\section{Environment}
\label{sec:environment}
%%%%%%%%%%%%%%%%%%%%%%%%%%%%
%%%%%%%%%%%%%%%%%%%%%%%%%%%%

There is a principal and $n$ agents. The principal chooses a physical outcome $x^i \in [0,1]$ and a payment $p^i \in \R$ for each agent $i \in \{1,\dots,n\}$.%
	\footnote{Nothing changes if $x^i$ lies in some compact and convex subset $\mathcal{X}^i$ of $\R$.}
We neglect for now any feasibility constraints that $(x^1,\dots,x^n)$ must satisfy.

Agent $i$'s payoff has the quasi-linear form $g^i(x^i,t^i) - p^i$,
where her (one-dimensional) type $t^i \in [0,1]$ is her private information.%
	\footnote{Again, nothing changes if $t^i$ lies in some compact and convex subset $\mathcal{T}^i$ of $\R$.}
The gross payoff function $g^i : [0,1] \times [0,1] \to \R$ satisfies two regularity conditions: the type derivative $g^i_2$ exists, and it is bounded.%
	\footnote{These conditions allow us to apply the envelope theorem of \textcite{MilgromSegal2002}.
	Boundedness can be weakened to the following: $g^i(x^i,\cdot)$ is absolutely continuous for each $x^i$, and $t^i \mapsto \sup_{x^i \in [0,1]} g^i_2(x^i,t^i)$ is dominated by an integrable function.}
Preferences are strictly single-crossing in the sense that $g^i_2(\cdot,t^i)$ is strictly increasing.

The leading example is a single-unit auction, in which physical outcomes are probabilities of obtaining the good. It is commonly assumed that gross payoffs have the product form $g^i(x^i,t^i) = x^i t^i$.

A direct mechanism is a map $(X,P) : [0,1]^n \to [0,1]^n \times \R^n$ specifying an outcome $X^i(t^i,t^{-i})$ and a payment $P^i(t^i,t^{-i})$ for each type $t^i \in [0,1]$ of each agent $i$, given the (reported) types $t^{-i}$ of the other agents. $(X,P)$ is \emph{weakly (strictly) strategy-proof} iff
\begin{align*}
	&g^i\left(X^i\left(t^i,t^{-i}\right),t^i\right) - P^i\left(t^i,t^{-i}\right) 
	\\
	\geq \mathrel{(>)}{} 
	&g^i\left(X^i\left(r^i,t^{-i}\right),t^i\right) - P^i\left(r^i,t^{-i}\right)
	\quad \text{for all $i$, $t^{-i}$ and $r^i \neq t^i$} .
\end{align*}
Strict strategy-proofness obviously implies weak strategy-proofness.

The following characterisation of weak strategy-proofness is well-known:%
	\footnote{See e.g. \textcite[Theorem 9.5]{JehleReny2011}. For this result, it suffices that preferences be \emph{weakly} single-crossing in the sense that $g^i_2(\cdot,t^i)$ is \emph{weakly} increasing.}
\begin{namedthm}[Spence--Mirrlees lemma.]
	A direct mechanism $(X,P)$ is weakly strategy-proof iff for each $i$ and $t^{-i}$, $X^i(\cdot,t^{-i})$ is weakly increasing and the envelope formula is satisfied:
	\begin{multline}
		P^i\left(t^i,t^{-i}\right) 
		= P^i\left(0,t^{-i}\right) 
		- g^i\left(X^i\left(0,t^{-i}\right),0\right)
		+ g^i\left(X^i\left(t^i,t^{-i}\right),t^i\right) 
		\\
		- \int_0^{t^i} g^i_2\left( X^i\left(s,t^{-i}\right), s \right) \dd s
		\quad \text{for all $t^i \in [0,1]$} .
		\label{eq:env}
		\tag{{\Large\Letter}}
	\end{multline}
\end{namedthm}

For strict strategy-proofness, we have the following natural analogue.

\begin{lemma}
	\label{lemma:strict_impl}
	A direct mechanism $(X,P)$ is \emph{strictly} strategy-proof iff for each $i$ and $t^{-i}$, $X^i(\cdot,t^{-i})$ is \emph{strictly} increasing and the envelope formula \eqref{eq:env} is satisfied.
\end{lemma}

The proof, given in the \hyperref[app]{appendix}, hews closely to a standard proof of the Spence--Mirrlees lemma.

\begin{remark}
	\label{remark:IR_LL}
	Suppose that each agent $i$ has a (type-independent) outside option worth $w^i$. Then strict strategy-proofness plus (ex post) individual rationality are equivalent to strict monotonicity, the envelope formula \eqref{eq:env}, and $g^i( X^i(0,t^{-i}), 0 ) - P^i(0,t^{-i}) \geq w^i$ for each $i$ and $t^{-i}$.
\end{remark}

\begin{remark}
	\label{remark:revelation_principle}
	Strict strategy-proofness does not admit a revelation principle:
	there are direct mechanisms $(X,P)$ which are not strictly strategy-proof,
	but which can be achieved as the strict dominant-strategy equilibrium of an \emph{indirect} mechanism.%
		\footnote{We are grateful to an anonymous referee for pointing this out.}
	For example, $(X,P) \equiv (0,0)$ is not strictly strategy-proof,
	but can be implemented in strictly dominant strategies (by a mechanism in which each agent has only one action available to her).
\end{remark}

%%%%%%%%%%%%%%%%%%%%%%%%%%%%
%%%%%%%%%%%%%%%%%%%%%%%%%%%%
\section{Strictness is essentially for free}
\label{sec:free}
%%%%%%%%%%%%%%%%%%%%%%%%%%%%
%%%%%%%%%%%%%%%%%%%%%%%%%%%%

Since any weakly increasing function is close to strictly increasing functions, \Cref{lemma:strict_impl} allows us to deduce that weak strategy-proofness can be made strict by an arbitrarily small modification.

We require a pair of additional assumptions: for each $i$, the gross payoff $g^i(\cdot,t^i)$ is continuous for each $t^i$, and the family $\{ g^i_2(\cdot,t^i) \}_{t^i \in [0,1]}$ of type derivatives is equi-continuous.%
	\footnote{$\{ g^i_2(\cdot,t^i) \}_{t^i \in [0,1]}$ is equi-continuous iff for any $\eps>0$, there is a $\delta>0$ such that $\abs{x^{i\prime}-x^i} < \delta$ implies that $\abs{ g^i_2(x^{i\prime},t^i) - g^i_2(x^i,t^i) } < \eps$ for every $t^i \in [0,1]$.}$^,$%
	\footnote{Continuity of $g^i(\cdot,t^i)$ for all $t^i$ may be weakened to continuity of $g^i(\cdot,0)$ only. But given the equi-continuity assumption, the latter implies the former, for $g^i(\cdot,t^i)$ equals $g^i(\cdot,0)$ plus a continuous integral (by Lebesgue's fundamental theorem of calculus).}
These assumptions are weak and typically satisfied in applications; for example, they hold for the gross payoff $g^i(x^i,t^i) = x^i t^i$ commonly used in auctions.

Let $\mathcal{X} \subseteq [0,1]^n$ be the set of feasible physical outcomes, and call a direct mechanism $(X,P)$ feasible iff $X(t)$ is feasible for every $t \in [0,1]^n$. For concreteness, assume that the feasible set is
\begin{equation*}
	\mathcal{X} 
	= \left\{ \left(x^1,\dots,x^n\right) \in [0,1]^n : 
	\sum_{i=1}^n x^i \leq 1 \right\} ,
\end{equation*}
which in the auction interpretation means that the good cannot be assigned with total probability exceeding unity. The argument below applies unchanged if the good must be assigned ($\sum_{i=1}^n x^i = 1$), or if there are no constraints ($\mathcal{X}=[0,1]^n$). Analogous arguments can be made for other constraints.

\begin{proposition}
	\label{proposition:denseness}
	If a feasible direct mechanism $(X,P)$ is weakly strategy-proof, then for any $\eps>0$, there is a feasible and strictly strategy-proof direct mechanism that is uniformly $\eps$-close to $(X,P)$.%
		\footnote{That is: for every $i$, $t^i$ and $t^{-i}$, $X^i(t^i,t^{-i})$ and $P^i(t^i,t^{-i})$ change by at most $\eps$.}
\end{proposition}

In the language of the virtual implementation literature (e.g. \textcite{AbreuSen1991,AbreuMatsushima1992}), \Cref{proposition:denseness} says that every feasible and weakly strategy-proof direct mechanism is \emph{virtually} strictly strategy-proof. Weak strategy-proofness therefore `essentially' shares the robustness properties of strict strategy-proofness; in particular, an arbitrarily small modification suffices to achieve full implementation.%
	\footnote{In the auction example, this modification adds some randomness to the allocation.}

\begin{proof}
	Let $(X,P)$ be feasible and weakly strategy-proof. We shall construct a feasible and strictly strategy-proof direct mechanism in terms of $\delta>0$, then show that we can make it uniformly $\eps$-close to $(X,P)$ by choosing $\delta$.

	For $\delta>0$, define a direct mechanism $(X_\delta,P_\delta)$ as follows. For each $i$,
	\begin{equation*}
		X_\delta^i\left(t^i,t^{-i}\right) 
		\coloneqq \delta \frac{ t^i + 1 }{ \sum_{j=1}^n \left( t^j + 1 \right) } 
		+ (1-\delta) X^i\left(t^i,t^{-i}\right) .
	\end{equation*}
	In the auction interpretation, this physical allocation rule assigns the good according to $X$ with probability $1-\delta$, and with probability $\delta$ assigns it randomly with probabilities strictly increasing in agents' types. This direct mechanism is feasible since $X$ is:
	\begin{equation*}
		\sum_{i=1}^n X_\delta^i\left(t^i,t^{-i}\right) 
		= \delta + (1-\delta) \sum_{i=1}^n X^i\left(t^i,t^{-i}\right)
		\leq 1 .
	\end{equation*}
	Since $(X,P)$ is weakly strategy-proof, $X^i(\cdot,t^{-i})$ is weakly increasing by the Spence--Mirrlees lemma; hence $X_\delta^i(\cdot,t^{-i})$ is strictly increasing.

	Define the payments to satisfy the envelope formula \eqref{eq:env}:
	\begin{multline*}
		P_\delta^i\left(t^i,t^{-i}\right)
		\coloneqq P^i\left(0,t^{-i}\right) 
		- g^i\left( X_\delta^i\left(0,t^{-i}\right), 0 \right) 
		+ g^i\left( X_\delta^i\left(t^i,t^{-i}\right), t^i \right) 
		\\
		- \int_0^{t^i} g^i_2\left( X_\delta^i\left(s,t^{-i}\right), s \right) \dd s .
	\end{multline*}
	Since strict monotonicity and the envelope formula hold, $(X_\delta,P_\delta)$ is strictly strategy-proof (for any $\delta>0$) by \Cref{lemma:strict_impl}.

	For every $i$, $t^i$ and $t^{-i}$, we have
	\begin{equation*}
		\abs*{ X_\delta^i\left( t^i, t^{-i} \right) - X^i\left( t^i, t^{-i} \right) }
		= \delta \abs*{ \frac{t^i+1}{\sum_{j=1}^n \left( t^j + 1 \right)} - X^i\left( t^i, t^{-i} \right) } .
	\end{equation*}
	Since $X^i$ maps into $[0,1]$, this expression is bounded by $\delta$, so that $X_\delta^i$ is uniformly $\delta$-close to $X^i$.

	Since $(X,P)$ is weakly strategy-proof, it satisfies \eqref{eq:env} by the Spence--Mirrlees lemma. Therefore, for every $i$, $t^i$ and $t^{-i}$,
	\begin{multline}
		\abs*{ P_\delta^i\left( t^i, t^{-i} \right) - P^i\left( t^i, t^{-i} \right) }
		\\
		\begin{aligned}[b]
			&=
			\bigg| 
			- \left[ g^i\left( X_\delta^i\left(0,t^{-i}\right), 0 \right) 
			- g^i\left( X^i\left(0,t^{-i}\right), 0 \right) \right]
			\\
			&\quad
			+ \left[ g^i\left( X_\delta^i\left(t^i,t^{-i}\right), t^i \right) 
			- g^i\left( X^i\left(t^i,t^{-i}\right), t^i \right) \right]
			\\
			&\quad
			- \int_0^{t^i} 
			\left[ g^i_2\left( X_\delta^i\left(s,t^{-i}\right), s \right) 
			- g^i_2\left( X^i\left(s,t^{-i}\right), s \right) \right]
			\dd s
			\bigg|
			\\
			&\leq
			\abs*{
			g^i\left( X_\delta^i\left(0,t^{-i}\right), 0 \right) 
			- g^i\left( X^i\left(0,t^{-i}\right), 0 \right) }
			\\
			&\quad
			+ \abs*{ g^i\left( X_\delta^i\left(t^i,t^{-i}\right), 0 \right) 
			- g^i\left( X^i\left(t^i,t^{-i}\right), 0 \right) }
			\\
			&\quad
			+ \int_0^{t^i} \abs*{ g^i_2\left( X_\delta^i\left(t^i,t^{-i}\right), s \right) 
			- g^i_2\left( X^i\left(t^i,t^{-i}\right), s \right) } \dd s
			\\
			&\quad
			+ \int_0^{t^i} 
			\abs*{ g^i_2\left( X_\delta^i\left(s,t^{-i}\right), s \right) 
			- g^i_2\left( X^i\left(s,t^{-i}\right), s \right) }
			\dd s ,
		\label{eq:P_diff}
		\end{aligned}
		\tag{$\vartriangle$}
	\end{multline}
	where we used Lebesgue's fundamental theorem of calculus.

	Now, fix $\eps > 0$. Further fix an agent $i$. Since $g^i(\cdot,0)$ is continuous, $\{ g^i_2(\cdot,s) \}_{s \in [0,1]}$ is equi-continuous, and $X_\delta^i$ is uniformly $\delta$-close to $X^i$ for any $\delta>0$, we may choose $\delta^i>0$ sufficiently small that the first two terms and the two integrands in \eqref{eq:P_diff} are all $<\eps/4$ for any $t^i$ and $t^{-i}$. Then
	\begin{equation*}
		\abs*{ P_{\delta^i}^i\left( t^i, t^{-i} \right) 
		- P^i\left( t^i, t^{-i} \right) }
		< \eps/4 + \eps/4
		+ \int_0^{t^i} \eps/4
		+ \int_0^{t^i} \eps/4
		\leq \eps 
	\end{equation*}
	for all $t^i$ and $t^{-i}$, which is to say that $P_{\delta^i}^i$ is uniformly $\eps$-close to $P^i$.

	Let $\delta \coloneqq \min\{ \eps, \delta^1, \dots, \delta^n \}$. Then for each $i$, $X_\delta^i$ is uniformly $\eps$-close to $X^i$, and $P_\delta^i$ is uniformly $\eps$-close to $P^i$.
\end{proof}

An implication of \Cref{proposition:denseness} is that the principal can render strategy-proofness strict at essentially no cost. Let
\begin{equation*}
	\boldsymbol{X} \times \boldsymbol{P} 
	= \left( [0,1]^n \right)^{[0,1]^n}
	\times \left( \R^n \right)^{[0,1]^n}
\end{equation*}
denote the set of all direct mechanisms.

\begin{corollary}
	\label{corollary:free}
	Let $(X,P)$ be feasible and weakly strategy-proof, and let $\Pi : \boldsymbol{X} \times \boldsymbol{P} \to \R$ be a continuous function.%
		\footnote{Continuity in the uniform topology is enough.}
	Then for any $\eps>0$, there is a feasible and strictly strategy-proof direct mechanism with value at least $\Pi(X,P) - \eps$.
\end{corollary}

The principal's objective function is typically continuous in applications. It often has the form
\begin{equation*}
	\Pi(X,P) = \int_{[0,1]^n} 
	\left[ \sum_{i=1}^n 
	\pi\left( X^i(t), P^i(t), t^i \right)
	\right]
	F\left( \dd t \right) ,
\end{equation*}
where $F$ is the principal's belief about how types are distributed. $\Pi$ is continuous for revenue-maximisation ($\pi(x,p,t) = p$), efficiency-maximisation ($\pi(x,p,t) = xt$), and agent-welfare-maximisation ($\pi(x,p,t) = xt - p$).

%%%%%%%%%%%%%%%%%%%%%%%%%%%%
%%%%%%%%%%%%%%%%%%%%%%%%%%%%
\section{Related literature}
\label{sec:related}
%%%%%%%%%%%%%%%%%%%%%%%%%%%%
%%%%%%%%%%%%%%%%%%%%%%%%%%%%

To our knowledge, strict strategy-proofness has appeared previously only in \textcite{BergemannMorris2009restud}.%
	\footnote{We thank Yi-Chun Chen for pointing out this reference.}
These authors characterise `robust' implementability by a direct mechanism in terms of strict ex-post incentive-compatibility (strict EPIC), which is equivalent to strict strategy-proofness in a private-values environment such as ours. In §7, they consider an auction environment, and construct a strictly EPIC direct mechanism close to an \emph{efficient} mechanism. \Cref{proposition:denseness} shows that with private values, this can be done for \emph{any} weakly strategy-proof mechanism, not only efficient ones. \textcite{BergemannMorris2009restud} do not characterise strict strategy-proofness, so have no analogue of \Cref{lemma:strict_impl}.

As mentioned in §\ref{sec:free}, \Cref{proposition:denseness} may be viewed as contributing to the virtual implementation literature (e.g. \textcite{AbreuSen1991,AbreuMatsushima1992}).
An analogous result is due to \textcite{BergemannMorris2009te}: they (we) show that virtual `robust' implementability (virtual strict strategy-proofness) is equivalent to weak strategy-proofness.%
	\footnote{This follows from their Theorem 2 and Corollary 1, specialised to our environment.
	Thus although strict strategy-proofness is more demanding than `robust' implementation, there is no gap between their virtual counterparts in our environment.}
This literature does not appear to have considered `strict' versions of incentive-compatibility.

More tangentially related are the literatures on undominated implementation (e.g. \textcite{Jackson1992}) and rationalisable implementation (e.g. \textcite{BergemannMorrisTercieux2011}). The former requires that agents' strategies not be weakly dominated; the latter that they not be iteratively strictly dominated. By contrast, we require agents' strategies to be strictly \emph{dominant}.

%______________________________________________________________________________

%       _                               _ _               
%      / \   _ __  _ __   ___ _ __   __| (_) ___ ___  ___ 
%     / _ \ | '_ \| '_ \ / _ \ '_ \ / _` | |/ __/ _ \/ __|
%    / ___ \| |_) | |_) |  __/ | | | (_| | | (_|  __/\__ \
%   /_/   \_\ .__/| .__/ \___|_| |_|\__,_|_|\___\___||___/
%           |_|   |_|                                     

% \pagebreak
\begin{appendices}
\crefalias{section}{appsec}
\crefalias{subsection}{appsec}
\crefalias{subsubsection}{appsec}

\renewcommand*{\thesubsection}{\Alph{subsection}}

%%%%%%%%%%%%%%%%%%%%%%
%%%%%%%%%%%%%%%%%%%%%%
\section*{Appendix: proof of \Cref{lemma:strict_impl}}
\label{app}
%%%%%%%%%%%%%%%%%%%%%%
%%%%%%%%%%%%%%%%%%%%%%

Begin with an observation: if the envelope formula \eqref{eq:env} holds, then the payoff loss of type $t^i$ of agent $i$ from mimicking type $r^i$, when the other agents' (reported) types are $t^{-i}$, is
\begin{multline}
	\left[ g^i\left(X^i\left(t^i,t^{-i}\right),t^i\right) 
	- P^i\left(t^i,t^{-i}\right) \right] 
	- \left[ g^i\left(X^i\left(r^i,t^{-i}\right),t^i\right) 
	- P^i\left(r^i,t^{-i}\right) \right] 
	\\
	= \int_{r^i}^{t^i} g^i_2\left(X^i\left(s,t^{-i}\right),s\right) \dd s
	- \left[ g^i\left(X^i\left(r^i,t^{-i}\right),t^i\right) 
	- g^i\left(X^i\left(r^i,t^{-i}\right),r^i\right) \right]
	\\
	= \int_{r^i}^{t^i} \left[ g^i_2\left(X^i\left(s,t^{-i}\right),s\right) 
	- g^i_2\left(X^i\left(r^i,t^{-i}\right),s\right) \right] \dd s .
	\label{eq:dev_payoff}
	\tag{$\star$}
\end{multline}
The final step used Lebesgue's fundamental theorem of calculus, which is applicable since $g^i_2$ is bounded.

Suppose that $(X,P)$ is strictly strategy-proof. Then it is weakly strategy-proof. Fix an $i$ and a $t^{-i}$. Since $g^i_2$ is bounded, the envelope theorem of \textcite[Theorem 2]{MilgromSegal2002} applies; hence the envelope formula \eqref{eq:env} must hold. Payoff losses from mimicking are therefore given by \eqref{eq:dev_payoff}. By strict strategy-proofness, \eqref{eq:dev_payoff} is strictly positive for any $r^i \neq t^i$. Since $g^i_2(\cdot,s)$ is increasing, this is possible only if $X^i(\cdot,t^{-i})$ is strictly increasing.%
	\footnote{In detail: consider $t^i>r^i$ (the case $t^i<r^i$ is analogous), and suppose toward a contradiction that $X^i(t^i,t^{-i}) \leq X^i(r^i,t^{-i})$. Since $g^i_2(\cdot,s)$ is increasing, we have
	\begin{equation*}
		\int_{r^i}^{t^i} g^i_2\left(X^i\left(t^i,t^{-i}\right),s\right) \dd s
		\leq \int_{r^i}^{t^i} g^i_2\left(X^i\left(r^i,t^{-i}\right),s\right) \dd s ,
	\end{equation*}
	which is equivalent to
	\begin{align*}
		&\int_{r^i}^{t^i} \left[ 
		g^i_2\left(X^i\left(s,t^{-i}\right),s\right) 
		- g^i_2\left(X^i\left(r^i,t^{-i}\right),s\right) 
		\right] \dd s
		\\
		+ &\int_{t^i}^{r^i} \left[ 
		g^i_2\left(X^i\left(s,t^{-i}\right),s\right) 
		- g^i_2\left(X^i\left(t^i,t^{-i}\right),s\right) 
		\right] \dd s
		\leq 0 .
	\end{align*}
	So it cannot be that both terms on the left-hand side are strictly positive. By \eqref{eq:dev_payoff}, this contradicts the hypothesis that $t^i$ strictly prefers not to mimic $r^i$ and vice-versa.}

Suppose that $(X,P)$ is such that for each $i$ and $t^{-i}$, $X^i(\cdot,t^{-i})$ is strictly increasing and the envelope formula \eqref{eq:env} holds. Then payoff losses from mimicking are given by \eqref{eq:dev_payoff}. By inspection, \eqref{eq:dev_payoff} is strictly positive for any $r^i \neq t^i$ since $g^i_2(\cdot,s)$ and $X^i(\cdot,t^{-i})$ are strictly increasing; hence $(X,P)$ is strictly strategy-proof. \qed

\end{appendices}

%______________________________________________________________________________

%    ____  _ _     _ _                             _           
%   | __ )(_) |__ | (_) ___   __ _ _ __ __ _ _ __ | |__  _   _ 
%   |  _ \| | '_ \| | |/ _ \ / _` | '__/ _` | '_ \| '_ \| | | |
%   | |_) | | |_) | | | (_) | (_| | | | (_| | |_) | | | | |_| |
%   |____/|_|_.__/|_|_|\___/ \__, |_|  \__,_| .__/|_| |_|\__, |
%                            |___/          |_|          |___/ 

% \pagebreak
\printbibliography[heading=bibintoc]

%______________________________________________________________________________

\end{document}